\documentclass{article}

\usepackage{arxiv}
\usepackage{cite}
\usepackage[utf8]{inputenc} 
\usepackage[T1]{fontenc}    
\usepackage{hyperref}       
\usepackage{url}            
\usepackage{booktabs}       
\usepackage{amsfonts}       
\usepackage{nicefrac}       
\usepackage{microtype}      
\usepackage{graphicx}
\usepackage{bm}
\usepackage{mathptmx}
\usepackage{dcolumn}

\title{Pseudo spin-valve switch based on ferromagnet/superconductor/ferromagnet trilayer microbridge.} 

\author{
    L.~N.~Karelina \\
Institute of Solid State Physics\\
Russian Academy of Sciences \\
Chernogolovka, 142432, Russia\\
\And
    V.~V.~Bolginov
\thanks{Also at Skobeltsyn Institute of Nuclear Physics, Lomonosov Moscow State University, Moscow 119991, Russia}\\
Institute of Solid State Physics\\
Russian Academy of Sciences\\
Chernogolovka, 142432, Russia\\
\And
    Sh.~A.~Erkenov
\thanks{Also at Institute of Solid State Physics, Russian Academy of Sciences, Chernogolovka, 142432, Russia}\\
Moscow Institute of Physics and Technology\\
State University\\
9 Institutskiy per., Dolgoprudny\\
Moscow Region, 141700, Russia\\
\And
    S.~V.~Egorov\\
Institute of Solid State Physics\\
Russian Academy of Sciences\\
Chernogolovka, 142432, Russia\\
\And
    I.~Golovchanskiy
\thanks{Also at Moscow Institute of Physics and Technology, State University, 9 Institutskiy per., Dolgoprudny, Moscow Region, 141700, Russia}\\
Russian National University of Science and Technology (NUST) MISiS\\
4 Leninsky Prospect, Moscow 119049, Russia\\
\And
    V.~I.~Chichkov\\
Russian National University of Science and Technology (NUST) MISiS\\
4 Leninsky Prospect, Moscow 119049, Russia\\
\And
    A.~ben~Hamida\\
Russian National University of Science and Technology (NUST) MISiS\\
4 Leninsky Prospect, Moscow 119049, Russia\\
\And
    V.~V.~Ryazanov
\footnotemark[3]\\
Institute of Solid State Physics\\
Russian Academy of Sciences\\
Chernogolovka, 142432, Russia
}

\begin{document}
\maketitle

\begin{abstract}
A noticeable magnetoresistive effect has been observed on
ferromagnet/superconductor/ferromagnet (FSF) microbridges based on
diluted ferromagnetic PdFe alloy containing as small as 1\% magnetic
atoms.  Microstructuring of the FSF trilayers does not destroy the
effect: the most pronounced curves were obtained on the smallest
bridges of \(6-8\ ~\mu \)m wide and \(10-15\ ~\mu \)m long. Below
the superconducting transition we are able to control the critical
current of microbridges by switching between P and AP orientations
of magnetizations of PdFe layers. The operation of FSF-bridge as a
magnetic switch is demonstrated in several regimes providing
significant voltage discrimination between digital states or
remarkably low bit error rate.
\end{abstract}


\section{Introduction}
At present, development and application of spin-valve devices based
on the giant magnetoresistance (GMR) is an intensive field of
science and technology (see, for example \cite{Tsymbal2001}). Use of
a superconducting interlayer (S) between two ferromagnetic metals
(F) instead of normal-metal interlayer was proposed in 1999
\cite{Tagirov1999, Buzdin1999}. ‘Pseudo spin-valve effect’ in
FSF-trilayers allows to control superconductivity of a thin
superconducting interlayer via mutual orientation of magnetizations
M$_1$ and M$_2$ in ferromagnetic layers \cite{Deutscher1969, Gu2002,
Moraru2006}. Several different manifestations of the spin-valve
effect are discussed. The most straightforward one is based on
suppression of superconductivity in $F_1$SF$_2$ heterostructures due
to spin-ordering antagonism between ferromagnetism and
superconductivity. In the case of co-directional M$_1$ and M$_2$
orientation (P-orientation) of the outer ferromagnetic layers the
superconducting layer critical temperature, $T_c$, is suppressed due
to proximity effect, while in the opposite case (AP-orientation) the
F-layer impacts slightly compensate each other and the suppression
is weakened \cite{Tagirov1999, Buzdin1999, Deutscher1969, Gu2002,
Moraru2006}. A similar effect is observed in the case when both
ferromagnetic layers are located on the same side of the
superconducting film (SFF structures) \cite{Oh1997}. In reality,
both the ‘positive’ effect with stronger suppression for
P-orientation and the ‘negative’ one with stronger suppression
for AP-orientation were observed experimentally depending on the
thickness of the ferromagnetic layers \cite{Leksin2012,
Fominov2010}. FSF and SFF-structures also demonstrate the triplet
spin-valve effect predicted in \cite{Fominov2010, Fominov2003,
Karminskaya2011} and observed in \cite{Zhu2010, Leksin2012a,
Zdravkov2013, Wang2014, Jara2014, Singh2015, Flokstra2015, Lenk2017}
in the case of non-collinear M$_1$ and M$_2$ mutual orientations.
The magnitude of the $T_c$ suppression due to the triplet spin-valve
effect ranges from 0.01~K up to 1.5~K depending on the spin-valve
geometry and materials of superconducting and ferromagnetic layers.

Usually the spin-valve effect is observed as peaks or dips of the
magnetoresistance at the coersive magnetic fields at which mutual
magnetizations of the ferromagnetic layers change. The positive
magnetoresistance (i.e. peaks of magnetoresistance) is observed in
SF$_1$F$_2$ structures due to leakage of the spin-triplet pairs from
superconducting layer into the ferromagnetic ones at non-collinear
magnetic configurations of the latter’s \cite{Zdravkov2013}. FSF
type samples usually demonstrate the negative magnetoresistance
\cite{Deutscher1969, Moraru2006, Zhu2010}. Positive
magnetoresistance of FSF structures could be caused by stray
magnetic fields of domain walls which arise in the coercive magnetic
field in the case of large samples \cite{Ryazanov2003, Rusanov2004,
Hwang2012}. To avoid this effect one of the ferromagnetic layers has
to be fixed by antifferomagnetic anchor layer (see, for example
\cite{Moraru2006}).

One of the noteworthy aspects of this work is the choice of the
material used for ferromagnetic layer deposition. The authors of
most previously published works used as the F-layers strong metallic
ferromagnets with in-plane magnetizations such as Fe, Ni,
Py\footnote[1]{Py is Fe$_{20}$Ni$_{80}$ or close alloy}, Co. Rare
earth ferromagnets (Ho, Dy) \cite{Gu2015} and half-mettalic CrO$_2$
\cite{Singh2015} were also applied to observe the spin-valve effect.
The weakest ferromagnet was Cu$_{1-x}$Ni$_x$ alloy (x $\approx$
50\%) with 40-70~K Curie temperature which was used in SFF and SFS
structures \cite{Gu2002, Zdravkov2013, Lenk2017}. Below we
demonstrate a noticeable magnetoresistance behavior using very
diluted Pd$_{0.99}$Fe$_{0.01}$ ferromagnet containing as small as
1\% at. magnetic atoms. Bulk Pd$_{1-x}$Fe$_x$ alloy is a
ferromagnetic material with long-range ferromagnetic order in the
range of Fe concentration $x = 0.001-1$ \cite{Heller1998,
Crangle1965}. Strongly diluted compositions with $x =
10^{-6}-10^{-2}$ undergo the ferromagnetic transition with Curie
temperature ranged from 10$^-4$ to 35~K \cite{Buscher1992,
Peters1984}. Polycrystalline Pd$_{1-x}$Fe$_x$ alloys remain
ferromagnetic down to the grain sizes of about 10~nm
\cite{Shinohara1999}. Thin films of the Pd$_{0.99}$Fe$_{0.01}$
content with the thickness below 100~nm demonstrate properties of a
nanocluster ferromagnet with weak interaction between nanoclusters
\cite{Uspenskaya2013, Golovchanskiy2016}. At the thickness below
25~nm the interaction between the clusters becomes drastically
weaker, because a three-dimensional distribution of the
ferromagnetic clusters arising around impurity iron atoms transforms
into a two-dimensional one \cite{Bolginov2017, Uspenskaya2014}. So
the spin-glass magnetic behavior starts to characterize the magnetic
response of the films to the external fields
\cite{Bolginov2017,Uspenskaya2017}. The
ferromagnetic-to-paramagnetic state transition occurs at the
thickness about of 10~nm \cite{Uspenskaya2014}. For practical
applications it means that thin Pd$_{0.99}$Fe$_{0.01}$ films possess
unique properties of soft magnetic material with weak magnetization
values, magnetization reversal in which occurs via independent
rotational processes in magnetic nanoclusters with the time $~ 3 - 5
\times 10^{-9}$~s, as it was shown by means of the FMR methods
\cite{Golovchanskiy2016}.

\begin{figure*}
\begin{center}
\includegraphics{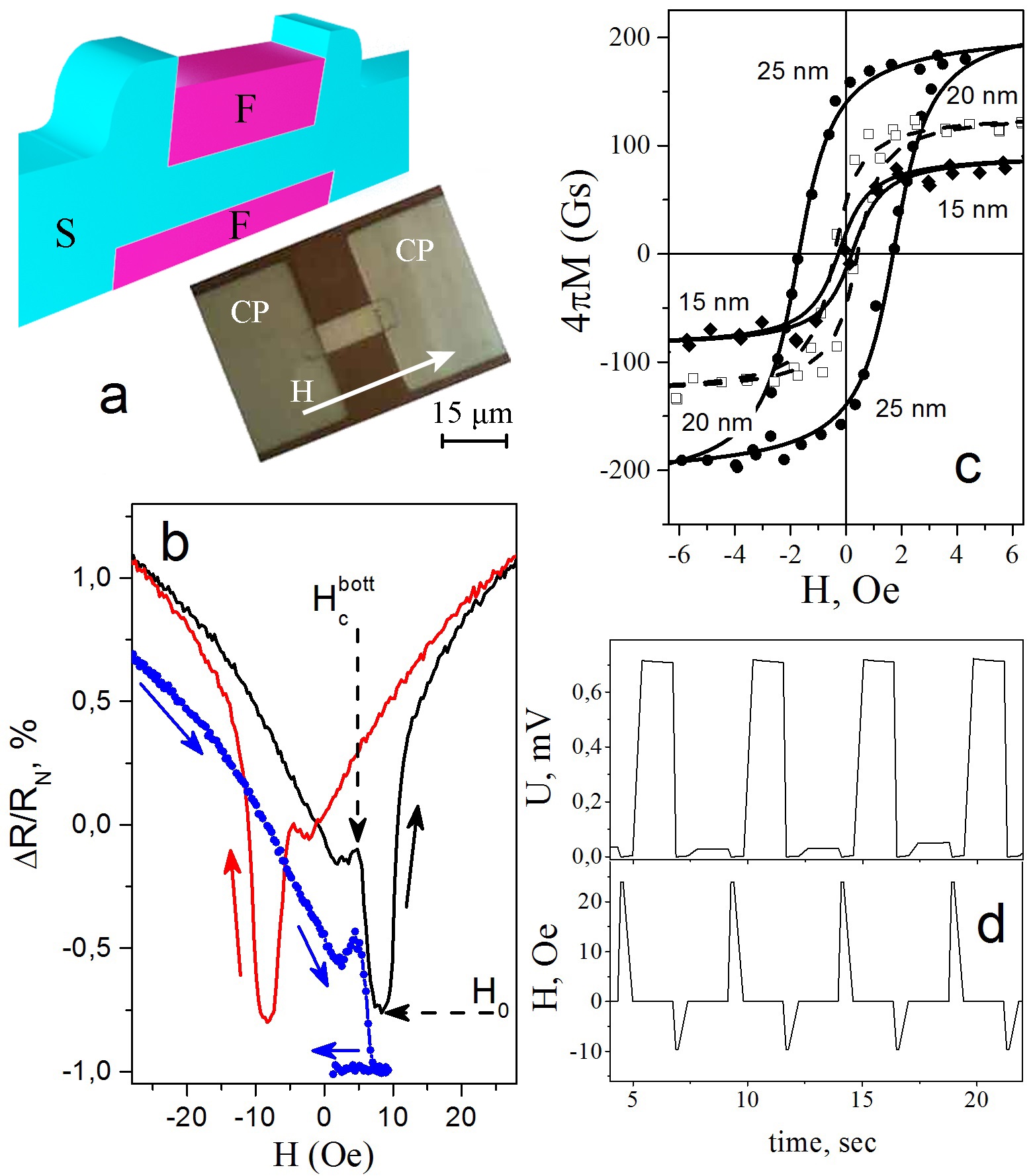}
\caption{\label{fig1}a)~A sketch of the FSF sample cross-section and
a microphotograph of \(8 \times 15 \mu\)m$^2$ FSF bridge located
between the niobium pads (CP). The arrow shows the direction of the
external magnetic field. The mark defines the scale in the
microphotograph. Definitions of $H_c^{bott}$ and $H_0$ are given in
the text.  b)~Magnetoresistance of the \(6 \times 15 \mu\)m$^2$
FSF-bridge measured in magnetic field swept from positive to
negative values and backward. Solid circles illustrate the magnetic
memory effect (see explanation in the text). The difference $R(H)$
to zero-field resistance $R(0)$ normalized on the normal resistance
$R_N$ at $T = $4.2~K is plotted. The temperature is 2.49~K (see
Fig.~\ref{fig3}). c)~Curves of magnetization reversal obtained for
\(10 \times 10 \mu\)m$^2$ PdFe layers of different thicknesses using
Josephson magnitometry method \cite{Bolginov2012} (see explanation
in the text). d)~The upper curve: PdFe-Nb-PdFe bridge switching
between two “logical states” due to external magnetic field
pulses shown below. The temperature is 2.217~K (see
Fig.~\ref{fig4}a). The bias current is \(57  \mu \)A. }
\end{center}
\end{figure*}

In this work we study the magnetoresistance of
Pd$_{0.99}$Fe$_{0.01}$-Nb-Pd$_{0.99}$Fe$_{0.01}$ trilayer structures
(FSF trilayers) in the form of microbridges of different width and
length (Fig.~\ref{fig1}a). As we will discuss below,
microstructuring of the SFS trilayers does not destroy the
magnetoresistive effects (Fig.~\ref{fig1}b) and allows achieving a
high supercurrent density as well as observing resistive switching
(Fig.~\ref{fig1}d) well below the critical temperature. Due to
reduction of the microbridge sizes we also achieve single-domain
\footnote[2]{more specifically, quasi-uniform} magnetic states of
ferromagnetic layers and eliminate the noted above effects of the
domain structure. The conventional pseudo spin-valve effect on the
resistive superconducting transition close to T$_c$ is also
investigated as a first step of our experimental study. Based on
results obtained in this work we have demonstrated the operation of
FSF microbridges as superconducting magnetic memory elements. Low
operation temperature allows to amplify additionally the switching
effect using non-linear shape of the current-voltage
characteristics.

\section{Samples and experiment}

\begin{figure*}
\begin{center}
\includegraphics{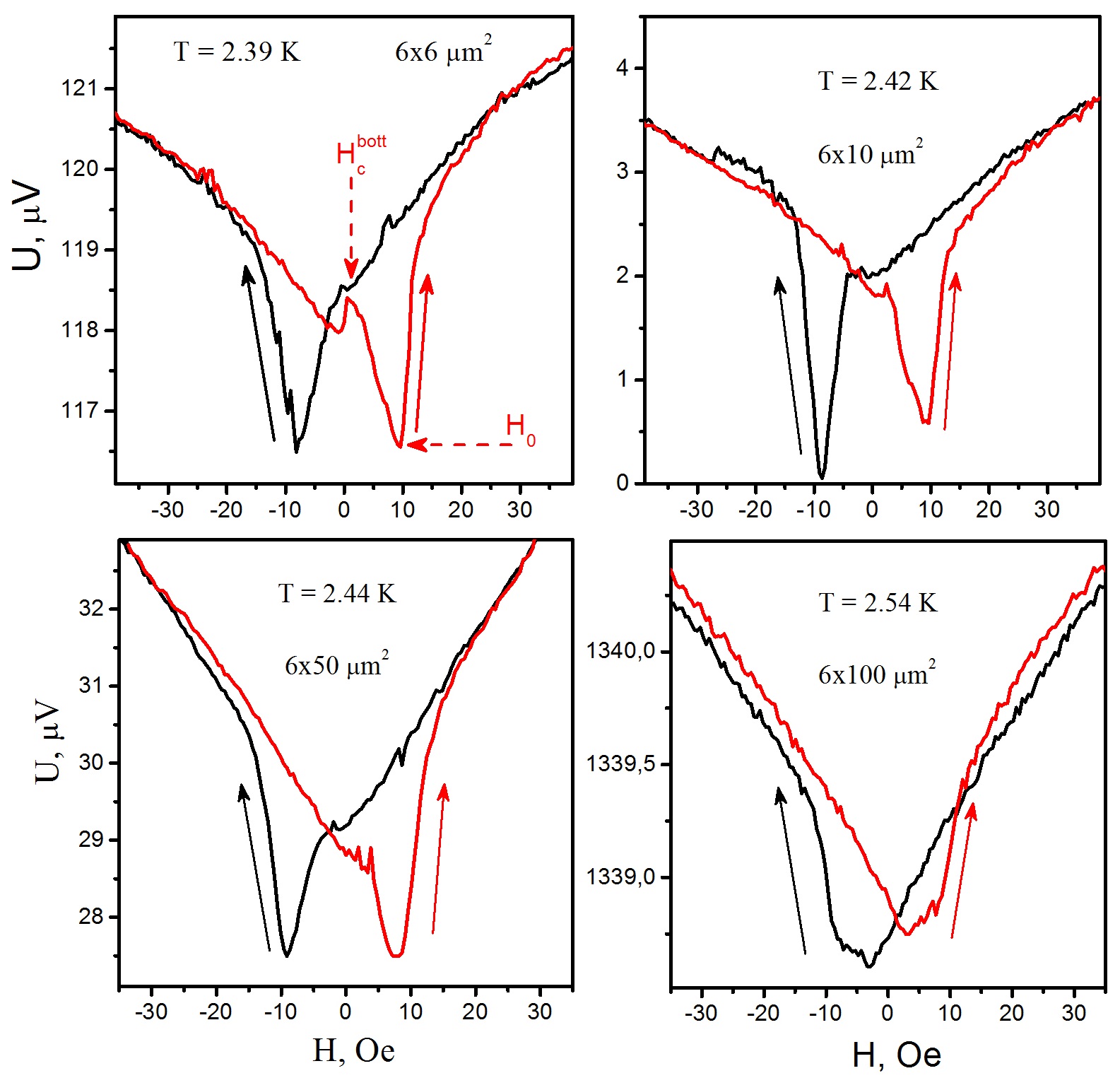}
\caption{\label{fig2}The magnetoresistance of FSF bridges with the
same F-layer width \(6 \mu \)m and different length and aspect
ratio. The length is represented for the top PdFe layer. The
thickness of the top PdFe layer is 45~nm and of the bottom PdFe
layer is 25~nm. Definitions of $H_c^{bott}$ and $H_0$ are given in
the text.  The bias current is \(20 \mu \)A.}
\end{center}
\end{figure*}

\begin{figure*}
\begin{center}
\includegraphics{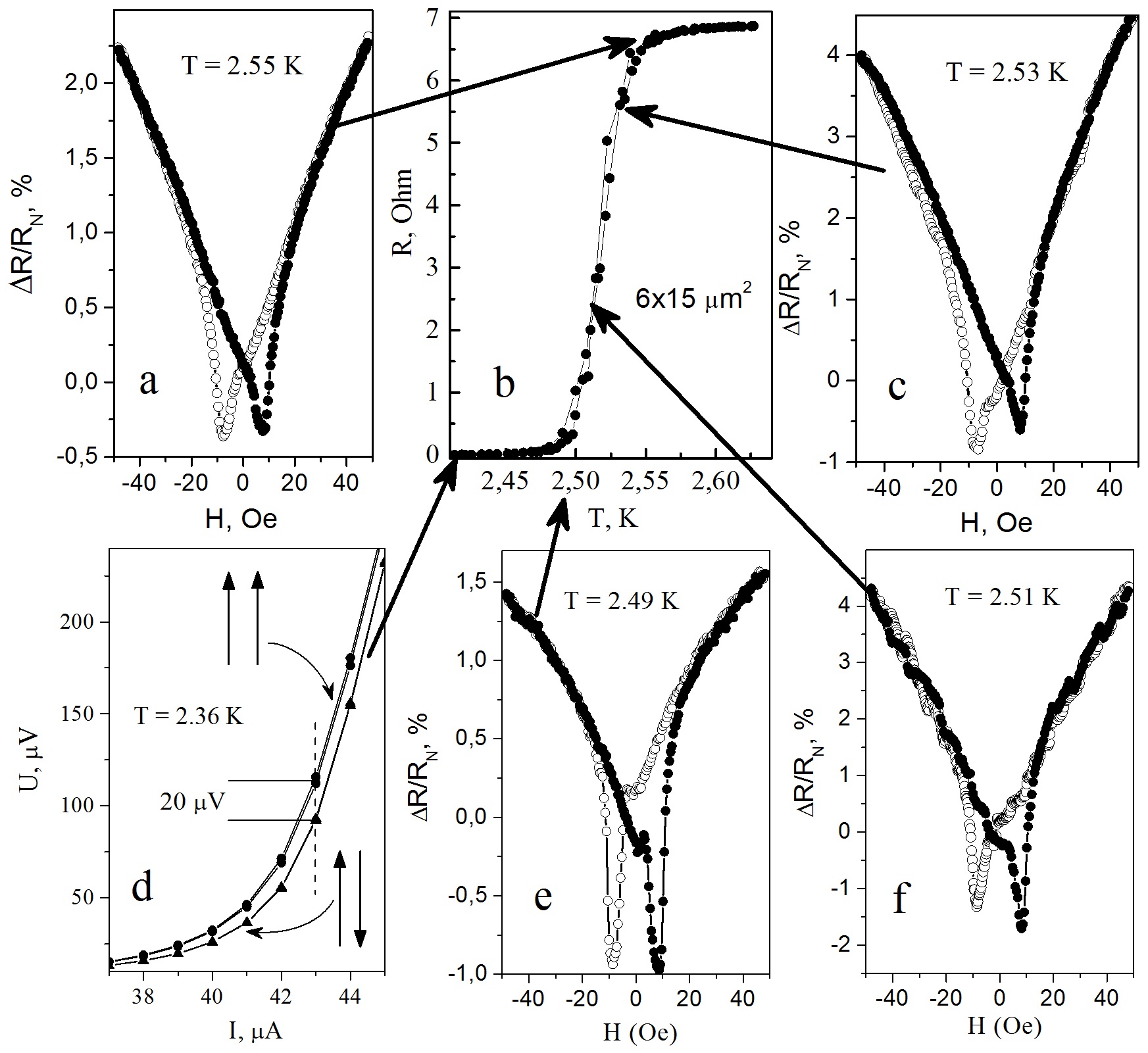}
\caption{\label{fig3}a,c,e,f)~Magnetoresistance obtained at
different temperatures within the superconducting transition for \(6
\times 15 \mu \)m$^2$ bridge. The difference $R(H)$ to zero-field
resistance $R(0)$ normalized on the normal resistance $R_N$ at $T =
$4.2~K is plotted. The bias current is \(20 \mu \)A.
b)~Superconducting transition of the \(6 \times 15 \mu \)m$^2$
sample. d)~ $IV$ curves for parallel and anti-parallel orientations
at $T = $2.36~K zoomed in near the critical current.}
\end{center}
\end{figure*}

The sample fabrication process starts from the PdFe-Nb-PdFe trilayer
deposition on Si/SiO$_2$ substrate by means of the rf-sputtering for
the Pd$_{0.99}$Fe$_{0.01}$ layers and the magnetron sputtering for
the Nb layer. We have studied several series of samples in which
thicknesses of top and bottom PdFe layers varied within 45-50~nm and
20-30~nm respectively (see Fig.~\ref{fig1}a). All samples showed
similar magnetoresistance with aspect ratio related features (see
Fig.~\ref{fig2} and a discussion below).  The thickness of Nb layer
was 15~nm and its critical temperature was strongly affected by the
proximity effect from adjacent PdFe layers. More precisely, the
critical temperature decreased from 7~K for single Nb film to
2.3-2.6~K for PdFe-Nb-PdFe trilayers while the measured width of the
superconducting transition was less than 0.05~K (see
Fig.~\ref{fig3}b). Then a series of FSF-rectangles of different
sizes were formed using the photolithography and argon ion milling.
The width $w$ of bridges ranged from \(4\ \mu \)m to \(25\ \mu \)m
while the length varied from \(14\ \mu \)m to \(108\ \mu \)m. At the
last stage Nb contact pads (CP) were fabricated using the magnetron
sputtering followed by the ‘lift-off’ process. The Nb layer was
of 120~nm thick and the contact pads were superconducting at all
temperatures below 8~K. To ensure good superconducting contact
between CP and the central superconducting layer the upper PdFe
layer was completely etched during the ion cleaning before the CP
deposition. So, the top PdFe layer was shorter by 8~nm due to
overlap of the contact pads and the bridge (see Fig.~\ref{fig1}a).
This value is taken below as a bridge length $L$.

Our earlier studies \cite{Bolginov2012} show that the Curie
temperature T$_{Curie}$ of Pd$_{0.99}$Fe$_{0.01}$ thin films
decreases with thickness due to the percolation nature of the
interaction between magnetic clusters \cite{Uspenskaya2013,
Bolginov2017, Uspenskaya2014}. In particular T$_{Curie} \approx$
13~K for 40~nm thick film and T$_{Curie}  \approx$ 9~K for 25~nm
thick film. The same is true for both the coercive fields and
saturation magnetization. This is confirmed by our research on SFS
and SIsFS Josephson junctions based on Pd$_{0.99}$Fe$_{0.01}$ alloy
by means of the Josephson magnetometry method (see
\cite{Bolginov2012, Larkin2012, Bakurskiy2013} and Fig.~\ref{fig1}c
with novel data). Thus, the coercive field for the bottom
ferromagnetic layer in the FSF bridges was smaller than for the top
one, and gradually reversing the magnetic field from large positive
value to negative one it was possible to achieve antiparallel
magnetization configuration in some range of negative magnetic
fields. In this field range one could expect the resistance
reduction due to decrease of the total spin-polarized electron
diffusion into the S-layer. Any antiferromagnetic anchor layers were
not used to fix the magnetization of one of the ferromagnetic layers
as it was done in earlier works \cite{Moraru2006, Leksin2012,
Zhu2010, Leksin2012a, Zdravkov2013} since it could enhance the
exchange interaction in PdFe layer. In addition, this is not
necessary in our case because the distorting stray-field effects
\cite{Ryazanov2003, Rusanov2004, Hwang2012} are not occurred when
using the weakly ferromagnetic PdFe alloy.

We started with the usual spin-valve effect study using
magnetoresistance measurements of FSF-bridges of different sizes at
various temperatures within the superconducting transition region.
Experiments were done in $^4$He cryostat equipped by a membrane
pressure stabilizer that allowed to fix the temperature with an
accuracy better than 0.01 K during the experiment. In
Fig.~\ref{fig1}b we present a typical magnetoresistive curve \(
\Delta\)$R(H)$, where \( \Delta\)$R=R(H)-R(0)$ and the magnetic
field $H$ is applied in plane parallel to the long side of the
bridge (see Fig.~\ref{fig1}a) by a superconducting solenoid. We
start from large positive magnetic field $H$, which is much larger
the saturation one for both F layers, and sweep $H$ to large
negative values and back. One can clearly see a sharp decrease of
the magnetoresistance at two magnetic fields $\pm H_0$ opposite in
sign to the initial saturated magnetizations. Let us consider in
more details the magnetization reversal as the external magnetic
field $H$ changes from large negative ($-$30~Oe) to large positive
($+$30~Oe) values. At $H = -$30~Oe both PdFe layers are magnetized
in the same (negative) direction. As $H$ reaches some small positive
value the thin (bottom) layer starts to turn its magnetization in
the positive direction. At first, the initial magnetic ordering
disappears completely at the coercive field for the bottom layer $H
= H_c^{bott}$ (see definition in Figs.~\ref{fig1}b,~\ref{fig2})
resulting in stray magnetic fields and positive magnetoresistance,
as discussed in \cite{Ryazanov2003, Hwang2012}. This is seen in
Fig.~\ref{fig1}b and Fig.~\ref{fig2} as a small rise in
magnetoresistance near $H_c^{bott}$. Further reorientation of local
magnetic moments in bottom F-layer in positive direction causes a
sharp decrease in the magnetoresistance. The main effect of the
negative magnetoresistance is obviously due to decrease of the total
spin-polarized electron diffusion into the S-layer at AP-orientation
of F-layers since this is the only possible mechanism mentioned in
introduction providing a negative magnetoresistance in
FSF-trilayers. As soon as the thick top F-layer starts to reverse
its magnetization (at $H \approx +H_0$ in Figs.~\ref{fig1}b
and~\ref{fig2}), the magnetoresistance begins to rise. In higher
fields, the magnetoresistance increases approximately as $H^2$ in
accordance with the Ginzburg-Landau theory (see, for example,
\cite{Schmidt1997}).

The next step was to study the effect of the bridge length and
aspect ratio on the magnetoresistive curves. The most pronounced
dips were observed for the smallest samples of \(4-20\ \mu \)m long
and \(6-8\ \mu \)m wide. As dimensions of the bridge increased these
dips became wider and gradually merged with the magnetoresistive
curve, appearing as small distortions. We believe this blurring of
the effect related to mutual F-layer magnetizations is due to domain
structure (more precisely – a long-range magnetic inhomogeneity
\cite{Uspenskaya2017, Uspenskaya2014}) of the long PdFe strips. This
is consistent with our earlier studies \cite{Bolginov2012} which
reveal a crossover from quasi-uniform to non-uniform magnetic state
as the PdFe film size increase from \(10 \times 10\ \mu \)m$^2$ to
\(30 \times 30\ \mu \)m$^2$.

The effect magnitude increases as the sample superconducting state
strengthens when temperature decrease at the region of the resistive
transition. In Fig.~\ref{fig3} we present the superconducting
resistive transition and a series of magnetoresistive curves
obtained for \(6 \times 15\ \mu \)m$^2$ bridge at various
temperatures within the resistive transition
$T_c^{(e)}<T<T_c^{(b)}$.\footnote[3]{Here $T_c^{e}=$2.55~K and
$T_c^{b}=$2.49~K denote the end and the beginning of the
superconducting transition and are introduced for clarity.} The
absolute magnitude of the effect increases from \(0.05~\Omega\) to
\(0.3~\Omega\), which corresponds to the resistance change up to 5\%
relative to the normal state resistance and up to 70\% relative to
the resistance at zero magnetic field at the corresponding
temperature. At the $T=$2.55~K $ \approx$ $T_c^{(b)}$ the dip depth
is about \(0.04~\Omega\) which is about 0.6\% to the normal state
resistance (Fig.~\ref{fig3}a). As the temperature decreases by
0.02~K the effect increases by 2.5~times (Fig.~\ref{fig3}c).  At $T
= $2.51~K two thirds of the resistive transition already passed and
the maximum dip depth of \(0.14~\Omega\)is achieved which
corresponds to 2\% of normal state resistance. At lowest temperature
$T = $2.49~K $ \approx$ $T_c^{(e)}$ the effect slightly decreases
since the bridge have a zero resistance in a vicinity of $H_0$. The
spin-valve effect was evaluated in terms of the critical temperature
variation in \cite{Deutscher1969, Gu2002, Moraru2006, Leksin2012,
Zhu2010, Leksin2012a, Zdravkov2013, Wang2014, Jara2014, Singh2015}.
We roughly estimate that the dip in magnetoresistive curve at $T =
$2.49~K is equivalent to the critical temperature change of about 1
mK. So small change is hardly possible to observe doing $R(T)$
measurement, but we were able to detect it on $R(H)$ curves in our
experimental situation. The relative effect magnitude normalized on
the zero-field resistance at given temperature rises from 0.6\% at
$T = T_c^{b}$ to 90\% at $T = T_c^{e}$. The temperature dependence
additionally confirms that the magnetoresistive effect is due to
enhance of the superconductivity of the niobium film rather than
PdFe anomalous magnetoresistance, for example.

It is important to note that investigated samples demonstrate a
magnetoresistive memory effect. In the experiments discussed above
the magnetic field is swept from large positive value to large
negative one and backward. In the measurement represented by the
blue line and symbols at the bottom of Fig.~\ref{fig1}b we have
reversed the backward sweep at $H_r = H_0$  and gradually decreased
magnetic field to zero. In this case the magnetoresistance remains
stable down to the negative field value corresponding to destruction
of the AP state. Note that the sweep can be reversed at any magnetic
field $H_r$ within the dip with some observed memory effect, however
the most pronounced effect takes place if $H_r = H_0$. This fact
additionally confirms that the effect is due to the steady AP
configuration of the ferromagnetic layer mutual magnetization and
this allows to use our FSF trilayers as superconducting logic
elements.

\begin{figure*}
\begin{center}
\includegraphics{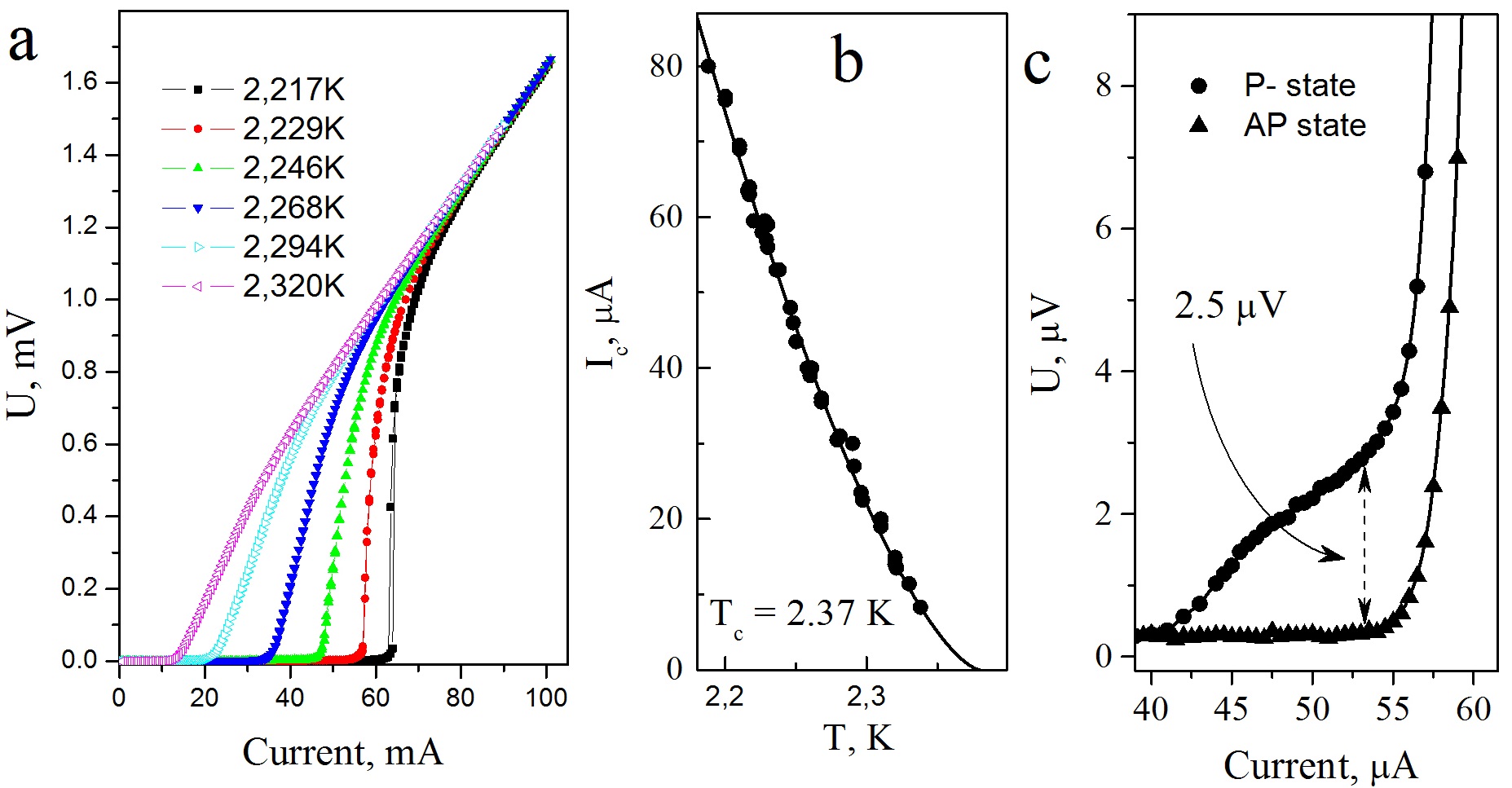}
\caption{\label{fig4}a)~The evolutuion of $IV$ curves with the
temperature in the superconducting states. b)~The temperature
dependence of the sample critical current. The line show
approximation be $(1-T/T_c)^{3/2}$ dependence.  c)~$IV$ curves for
parallel and anti-parallel orientations at $T = $2.217~K zoomed in
near the critical current. }
\end{center}
\end{figure*}

Below the superconducting transition ($T<T_c^{e}$), the
current-voltage characteristics of the microbridges cease to be
linear and acquire a region of zero resistance at $I < I_c$
(Fig.~\ref{fig4}a). Magnetoresistive curves similar to those shown
in Figs.~\ref{fig3}a,c,e,f can also be observed at these
temperatures if the bias current is slightly above $I_c$. We also
found that below $T_c^{e}$ the effect of mutual magnetization of
ferromagnetic layers is to be described in terms of the critical
current change (Fig.~\ref{fig3}d). The ability to control the
critical current expands the working temperature range of the
spin-valves based on the FSF-microbridges and gives the advantage
compared to conventional FSF spin-valve structures
\cite{Deutscher1969, Gu2002,Moraru2006, Oh1997, Leksin2012,
Fominov2010, Fominov2003, Karminskaya2011,Zhu2010, Leksin2012a,
Zdravkov2013, Wang2014, Jara2014, Singh2015, Flokstra2015, Lenk2017}
operating in a narrow temperature range near the superconducting
transition. The temperature dependence of the critical current for
SFS bridge with AP mutual F-layer magnetizations obeys the
well-known $(1-T/T_c)^{3/2}$ low for thin film depairing current
(see Fig.~\ref{fig4}b). A shape of the $IV$ characteristic
transforms steadily from a linear curve at $T=T_c^{e}$ to highly
nonlinear one with a zero-resistance region and a sharp hopping into
resistive state similar to that in \cite{Rusanov2004a}. In
Fig.~\ref{fig4}a the latter occurs at $T = 0.93T_c$ ($T = $2.2~K),
where $T_c \approx $2.37~K was determined as the fitting parameter
in Fig.~\ref{fig4}b. At lower temperatures $IV$ curves demonstrate
hysteretic behavior due to sample overheating in the resistive
state. Moreover, the hopping is driven mainly by noise and a
reproducible switching between the normal and the superconducting
states is not possible. At intermediate temperatures $IV-$curves for
both P and AP orientations are single-valued and one can control the
voltage on the sample via the mutual orientation of PdFe
magnetizations.

We are interested in maximal voltage difference between AP and P
‘digital states’, which makes it easier to detect logical
switching of the memory element, increases the corresponding
Josephson frequency (i.e. readout rate) and determines a possible
integration the element into superconducting digital RSFQ-circuitry.
This difference increases at lower temperatures due to increasing
non-linearity of $IV-$curves (Fig.~\ref{fig4}a). The maximum voltage
difference about \(600 \mu \)V (Fig.~\ref{fig1}d) was obtained close
to the boundary temperature (at $T = $2.217~K in Fig.~\ref{fig4}a).
At this temperature range on the $IV$ curve a resistive region
appears for P state in which  the voltage changes quite smoothly
with the current in the same range, where the AP state is
superconducting (Fig.~\ref{fig4}c). For the bias current marked with
an arrow in the Fig.~\ref{fig4}c we can obtain a reproducible
switching between \(2.5 \mu \)V and, in fact, the superconducting
state ($U = 0$). In other words, the magnetoresistance ratio
$(R_P-R_{AP})/ R_{AP}$ is infinitely large. In Fig.~\ref{fig1}d
switches between \(700 \mu \)V and \(25 \mu \)V voltage levels are
represented with magnetoresistance ratio equal 27.  The low voltage
level in the resistive state of the FSF bridges eliminates
overheating and completely avoids accidental switching. The width of
the low-voltage region, which reaches 30\% of the critical current,
provides large bias current margins. The origin of the smooth
low-voltage region in not entirely clear now but it is undoubtedly
very important for applications.

\section{Discussion}

In this paper we propose new FSF memory elements in the form of
microbridges and we would like to discuss their advantages and
disadvantages compared to unstructured FSF spin-valve devices. The
result shown in Fig.~\ref{fig1}d and Fig.~\ref{fig4}c reliably
proves that the memory effect can manifest itself in FSF
microbridges noticeably below $T_c$ and use of the PdFe-Nb-PdFe
microbridges as memory elements is still possible at low
temperatures, not just in the superconducting transition region. The
idea to control the critical current rather than the critical
temperature was proposed firstly in the work by A. Rusanov et al
\cite{Rusanov2004a}. This approach distinguishes \cite{Rusanov2004a}
and this work from the most of previous ones in which FSF trilayers
have not been structured and only measurements in the range of the
resistive transition were possible. Surprisingly we have found only
few works studying superconductor-ferromagnet microbridges
\cite{Ryazanov2003, Rusanov2004, Rusanov2004a} and demonstrating
positive magnitoresistive effects related to remagnetization of
F-layers. The use of a very weak ferromagnet makes it possible to
observe a distinct negative magnetoresistance effect, which is not
distorted by the magnetostatic interaction of the layers during the
sample magnetization reversal. It was shown in \cite{Moraru2006,
Leksin2012, Zhu2010, Leksin2012a, Zdravkov2013} that an anchor
anti-ferromagnetic layer is a key element to observe the spin-valve
effect using traditional ferromagnets (for example, Py). The cluster
nature of the PdFe magnetism allows the magnetic field lines to
close inside the ferromagnetic film, which reduces the domain
formation, weakens stray fields and the magnetostatic interaction as
a whole. The use of the “cryogenic” ferromagnet also provided a
fairly wide transition area between the resistive and
superconducting regions, which allowed us to study the evolution of
the shape of current-voltage characteristics, to find conditions for
zero resistance in a low-voltage state and maximum voltage amplitude
of magnetic switching.

The results obtained are important primarily for development of the
Josephson magnetic memory. In earlier works of our group
\cite{Bolginov2012, Larkin2012, Bakurskiy2013} digital states of
Josephson memory elements were determined by values of the
magnetization flux through the cross-section of the multilayered SFS
or SIsFS junctions, which resulted from the remagnetization of a
single PdFe thin layer. This type of Josephson memory elements is
hardly to be scalable since the magnetic flux vanishes with the
junction sizes. On the contrary, FSF bridges can decrease almost
unlimitedly, down to deeply submicron-scale sizes. The enhancement
of the detected memory effect can be achieved using PdFe-Nb-PdFe
heterostructures as a Josephson barrier in multilayer Josephson
S-(F1sF2)-S junctions \cite{Klenov2019}. The high transparency of
the diluted Pd$_{0.99}$Fe$_{0.01}$ alloy for the Josephson
supercurrent is an important advantage for use in these devices.

To conclude, in the present paper magnetoresistive characteristics
and switches between superconducting and resistive state of trilayer
FSF (Pd$_{0.99}$Fe$_{0.01}$-Nb-Pd$_{0.99}$Fe$_{0.01}$) bridges were
studied. Despite of very low content of magnetic atoms in
Pd$_{0.99}$Fe$_{0.01}$ resulting in low Curie temperature we have
observed noticeable negative magnetoresistance at antiparallel
orientation of magnetizations of PdFe layers in the FSF structures.
At temperatures noticeably lower than the resistive transition to
the superconducting state the change of the F-layers mutual
magnetization allows to control the critical current of the SFS
microbridges. This provides a demonstration of the magnetic memory
effect with switching between superconducting (or low-voltage) state
and resistive state with the voltage change up to 0.6~mV. An
important advantage of FSF microbridges in comparison with Josephson
SFS memory previously proposed is the ability to reduce these
structures down to deeply submicron-scale sizes.


%

The work is published under support of RFBR grant No.~19--32--90162.

\bibliographystyle{unsrt}


\begin{thebibliography}{10}

\bibitem{Tsymbal2001}
E.~Y. Tsymbal and D.~G. Pettifor.
\newblock {Perspectives of giant magnetoresistance}.
\newblock In H.~Ehrenreich and F.~Spaepen, editors, {\em Solid State Physics -
  Advances in Research and Applications}, volume~56, pages 113--237. Academic
  Press, 2001.

\bibitem{Tagirov1999}
L.~R. Tagirov.
\newblock {Low-field superconducting spin switch based on a
  superconductor/ferromagnet multilayer}.
\newblock {\em Physical Review Letters}, 83:2058--2061, 1999.

\bibitem{Buzdin1999}
A.~I. Buzdin, A.~V. Vedyayev, and N.~V. Ryzhanova.
\newblock {Spin-orientation – dependent superconductivity in F / S / F
  structures}.
\newblock {\em Europhysics Letters}, 48:686, 1999.

\bibitem{Deutscher1969}
G.~Deutscher and F.~Meunier.
\newblock {Coupling Between Ferromafnetic Layers Through a Superconductor}.
\newblock {\em Physical Review Letters}, 22(9):395--396, 1969.

\bibitem{Gu2002}
J.~Y. Gu, Chun~Yeol You, J.~S. Jiang, and S.~D. Bader.
\newblock {Magnetization-orientation dependence of the superconducting
  transition temperature and magnetoresistance in the
  ferromagnet-superconductor-ferromagnet trilayer system: CuNi/Nb/CuNi}.
\newblock {\em Physical Review Letters}, 89:267001, 2002.

\bibitem{Moraru2006}
Ion~C. Moraru, W.~P. Pratt, and Norman~O. Birge.
\newblock {Magnetization-dependent T$_c$ shift in
  ferromagnet/superconductor/ferromagnet trilayers with a strong ferromagnet}.
\newblock {\em Physical Review Letters}, 96:037004, 2006.

\bibitem{Oh1997}
Sangjun Oh, D.~Youm, and M.~R. Beasley.
\newblock {A superconductive magnetoresistive memory element using controlled
  exchange interaction}.
\newblock {\em Applied Physics Letters}, 71:2376--2378, 1997.

\bibitem{Leksin2012}
P.~V. Leksin, N.~N. Garif'Yanov, I.~A. Garifullin, J.~Schumann, V.~Kataev,
  O.~G. Schmidt, and B.~B{\"{u}}chner.
\newblock {Physical properties of the superconducting spin-valve Fe/Cu/Fe/In
  heterostructure}.
\newblock {\em Physical Review B}, 85:024502, 2012.

\bibitem{Fominov2010}
Ya.~V. Fominov, A.~A. Golubov, T.~Yu. Karminskaya, M.~Yu. Kupriyanov, R.~G.
  Deminov, and L.~R. Tagirov.
\newblock {Superconducting triplet spin valve}.
\newblock {\em JEPT Lett.}, 91:308--313, 2010.

\bibitem{Fominov2003}
Ya~V. Fominov, A.~A. Golubov, and M.~Yu Kupriyanov.
\newblock {Triplet proximity effect in FSF trilayers}.
\newblock {\em Journal of Experimental and Theoretical Physics Letters},
  77:510--515, 2003.

\bibitem{Karminskaya2011}
T.~Yu Karminskaya, A.~A. Golubov, and M.~Yu Kupriyanov.
\newblock {Anomalous proximity effect in spin-valve
  superconductor/ferromagnetic metal/ferromagnetic metal structures}.
\newblock {\em Physical Review B}, 84:064531, 2011.

\bibitem{Zhu2010}
Jian Zhu, Ilya~N. Krivorotov, Klaus Halterman, and Oriol~T. Valls.
\newblock {Angular dependence of the superconducting transition temperature in
  ferromagnet-superconductor-ferromagnet trilayers}.
\newblock {\em Physical Review Letters}, 105:207002, 2010.

\bibitem{Leksin2012a}
P.~V. Leksin, N.~N. Garif'Yanov, I.~A. Garifullin, Ya~V. Fominov, J.~Schumann,
  Y.~Krupskaya, V.~Kataev, O.~G. Schmidt, and B.~B{\"{u}}chner.
\newblock {Evidence for triplet superconductivity in a
  superconductor-ferromagnet spin valve}.
\newblock {\em Physical Review Letters}, 109:057005, 2012.

\bibitem{Zdravkov2013}
V.~I. Zdravkov, J.~Kehrle, G.~Obermeier, D.~Lenk, H.~A. {Krug Von Nidda},
  C.~M{\"{u}}ller, M.~Yu Kupriyanov, A.~S. Sidorenko, S.~Horn, R.~Tidecks, and
  L.~R. Tagirov.
\newblock {Experimental observation of the triplet spin-valve effect in a
  superconductor-ferromagnet heterostructure}.
\newblock {\em Physical Review B}, 87:144507, 2013.

\bibitem{Wang2014}
X.~L. Wang, A.~{Di Bernardo}, N.~Banerjee, A.~Wells, F.~S. Bergeret, M.~G.
  Blamire, and J.~W.A. Robinson.
\newblock {Giant triplet proximity effect in superconducting pseudo spin valves
  with engineered anisotropy}.
\newblock {\em Physical Review B}, 89:140508, 2014.

\bibitem{Jara2014}
Alejandro~A. Jara, Christopher Safranski, Ilya~N. Krivorotov, Chien~Te Wu,
  Abdul~N. Malmi-Kakkada, Oriol~T. Valls, and Klaus Halterman.
\newblock {Angular dependence of superconductivity in superconductor/spin-valve
  heterostructures}.
\newblock {\em Physical Review B}, 89:184502, 2014.

\bibitem{Singh2015}
A.~Singh, S.~Voltan, K.~Lahabi, and J.~Aarts.
\newblock {Colossal proximity effect in a superconducting triplet spin valve
  based on the half-metallic ferromagnet CrO$_2$}.
\newblock {\em Physical Review X}, 5:021019, 2015.

\bibitem{Flokstra2015}
M.~G. Flokstra, T.~C. Cunningham, J.~Kim, N.~Satchell, G.~Burnell, P.~J.
  Curran, S.~J. Bending, C.~J. Kinane, J.~F.K. Cooper, S.~Langridge,
  A.~Isidori, N.~Pugach, M.~Eschrig, and S.~L. Lee.
\newblock {Controlled suppression of superconductivity by the generation of
  polarized Cooper pairs in spin-valve structures}.
\newblock {\em Physical Review B}, 91:060501, 2015.

\bibitem{Lenk2017}
D.~Lenk, R.~Morari, V.~I. Zdravkov, A.~Ullrich, Yu~Khaydukov, G.~Obermeier,
  C.~M{\"{u}}ller, A.~S. Sidorenko, H.~A.Krug {Von Nidda}, S.~Horn, L.~R.
  Tagirov, and R.~Tidecks.
\newblock {Full-switching FSF-type superconducting spin-triplet magnetic random
  access memory element}.
\newblock {\em Physical Review B}, 96:184521, 2017.

\bibitem{Ryazanov2003}
V~V Ryazanov, V~A Oboznov, A~S Prokof'ev, and S~V Dubonos.
\newblock {Proximity effect and spontaneous vortex phase in planar SF
  structures}.
\newblock {\em Journal of Experimental and Theoretical Physics Letters},
  77:39--43, 2003.

\bibitem{Rusanov2004}
A.~Yu Rusanov, M.~Hesselberth, J.~Aarts, and A.~I. Buzdin.
\newblock {Enhancement of the superconducting transition temperature in
  Nb/permalloy bilayers by controlling the domain state of the ferromagnet}.
\newblock {\em Physical Review Letters}, 93:057002, 2004.

\bibitem{Hwang2012}
Tae-Jong Hwang and Dong~Ho Kim.
\newblock {Influence of stray fields and the proximity effect in
  ferromagnet/superconductor/ferromagnet spin valves}.
\newblock {\em Journal of the Korean Physical Society}, 61:1628--1632, 2012.

\bibitem{Gu2015}
Yuanzhou Gu, G{\'{a}}borb Hal{\'{a}}sz, J.~W.A. Robinson, and M.~G. Blamire.
\newblock {Large Superconducting Spin Valve Effect and Ultrasmall Exchange
  Splitting in Epitaxial Rare-Earth-Niobium Trilayers}.
\newblock {\em Physical Review Letters}, 115:067201, 2015.

\bibitem{Heller1998}
B.~Heller, K.~H. Speidel, R.~Ernst, A.~Gohla, U.~Grabowy, V.~Roth, G.~Jakob,
  F.~Hagelberg, J.~Gerber, S.~N. Mishra, and P.~N. Tandon.
\newblock {Transient field measurement in the giant moment PdFe alloy}.
\newblock {\em Nuclear Instruments and Methods in Physics Research B},
  142:133--138, 1998.

\bibitem{Crangle1965}
J.~Crangle and W.~R. Scott.
\newblock {Dilute ferromagnetic alloys}.
\newblock {\em Journal of Applied Physics}, 36:921--928, 1965.

\bibitem{Buscher1992}
C.~B{\"{u}}scher, T.~Auerswald, E.~Scheer, A.~Schr{\"{o}}der, H.~V.
  L{\"{o}}hneysen, and H.~Claus.
\newblock {Ferromagnetic transition in dilute Pd-Fe alloys}.
\newblock {\em Physical Review B}, 46:983--989, 1992.

\bibitem{Peters1984}
R~P Peters, Ch. Buchal, M~Kubota, R~M Mueller, and F~Pobell.
\newblock {Palladium-Iron: A Giant-Moment Spin-Glass at Ultralow Temperatures}.
\newblock {\em Physical Review Letters}, 53(11):1108--1111, 1984.

\bibitem{Shinohara1999}
T~Shinohara, T~Sato, T~Taniyama, and I~Nakatani.
\newblock {Size dependent magnetization of PdFe fine particles}.
\newblock {\em Journal of Magnetism and Magnetic Materials}, 196-197:94--95,
  1999.

\bibitem{Uspenskaya2013}
L.~S. Uspenskaya, A.~L. Rakhmanov, L.~A. Dorosinskii, A.~A. Chugunov, V.~S.
  Stolyarov, O.~V. Skryabina, and S.~V. Egorov.
\newblock {Magnetic patterns and flux pinning in Pd$_{0.99}$Fe$_{0.01}$-Nb
  hybrid structures}.
\newblock {\em JETP Letters}, 97:155--158, 2013.

\bibitem{Golovchanskiy2016}
I.~A. Golovchanskiy, V.~V. Bolginov, N.~N. Abramov, V.~S. Stolyarov, A.~{Ben
  Hamida}, V.~I. Chichkov, D.~Roditchev, and V.~V. Ryazanov.
\newblock {Magnetization dynamics in dilute Pd$_{1-x}$Fe$_x$ thin films and
  patterned microstructures considered for superconducting electronics}.
\newblock {\em Journal of Applied Physics}, 120:163902, 2016.

\bibitem{Bolginov2017}
V.~V. Bol'ginov, O.~A. Tikhomirov, and L.~S. Uspenskaya.
\newblock {Two-component magnetization in Pd$_{0.99}$Fe$_{0.01}$ thin films}.
\newblock {\em JETP Letters}, 105:169--173, 2017.

\bibitem{Uspenskaya2014}
L~S Uspenskaya, A~L Rakhmanov, L~A Dorosinskii, S~I Bozhko, V~S Stolyarov, and
  V~V Bolginov.
\newblock {Magnetism of ultrathin Pd$_{0.99}$Fe$_{0.01}$ films grown on
  niobium}.
\newblock {\em Materials Research Express}, 1:036104, 2014.

\bibitem{Uspenskaya2017}
L.~S. Uspenskaya and I.~N. Khlyustikov.
\newblock {Anomalous magnetic relaxation in thin Pd$_{0.99}$Fe$_{0.01}$ films}.
\newblock {\em Journal of Experimental and Theoretical Physics}, 125:875--878,
  2017.

\bibitem{Bolginov2012}
V.~V. Bol'ginov, V.~S. Stolyarov, D.~S. Sobanin, A.~L. Karpovich, and V.~V.
  Ryazanov.
\newblock {Magnetic switches based on Nb-PdFe-Nb Josephson junctions with a
  magnetically soft ferromagnetic interlayer}.
\newblock {\em JETP Letters}, 95:366--371, 2012.

\bibitem{Larkin2012}
Timofei~I. Larkin, Vitaly~V. Bol'ginov, Vasily~S. Stolyarov, Valery~V.
  Ryazanov, Igor~V. Vernik, Sergey~K. Tolpygo, and Oleg~A. Mukhanov.
\newblock {Ferromagnetic Josephson switching device with high characteristic
  voltage}.
\newblock {\em Applied Physics Letters}, 100:222601, 2012.

\bibitem{Bakurskiy2013}
S.~V. Bakurskiy, N.~V. Klenov, I.~I. Soloviev, V.~V. Bolginov, V.~V. Ryazanov,
  I.~V. Vernik, O.~A. Mukhanov, M.~Yu Kupriyanov, and A.~A. Golubov.
\newblock {Theoretical model of superconducting spintronic SIsFS devices}.
\newblock {\em Applied Physics Letters}, 102(19):192603, 2013.

\bibitem{Schmidt1997}
V.~V. Schmidt.
\newblock {\em {The Physics of Superconductors (Section 3.5)}}.
\newblock Springer-Verlag, 1997.

\bibitem{Rusanov2004a}
A.~Rusanov, M.~Hesselberth, S.~Habraken, and J.~Aarts.
\newblock {Depairing currents in superconductor ferromagnet Nb/CuNi trilayers
  close to Tc}.
\newblock {\em Physica C}, 404:322--325, 2004.

\bibitem{Klenov2019}
Nikolay Klenov, Yury Khaydukov, Sergey Bakurskiy, Roman Morari, Igor Soloviev,
  Vladimir Boian, Thomas Keller, Mikhail Kupriyanov, Anatoli Sidorenko, and
  Bernhard Keimer.
\newblock {Periodic Co / Nb pseudo spin valve for cryogenic memory}.
\newblock {\em Beilstein Journal of Nanotechnology}, 10:833--839, 2019.

\end{thebibliography}

\end{document}